\documentclass[twocolumn, prd, aps, longbibliography]{revtex4-2}
\usepackage{graphicx}
\usepackage{hyperref}
\usepackage{amssymb}
\usepackage{amsmath}
\usepackage{xcolor}
\usepackage{lineno}
\usepackage{enumitem}
\usepackage{xspace}
\newcommand{\grb}{GRB~221009A\xspace}
\newcommand{\dd}{\text{d}}
\newcommand{\gammabg}{\gamma_\text{bg}}
\newcommand{\nucleus}[2]{^{#1}_{#2} \text{X}}

\begin{document}

\title{\grb: a potential source of ultra-high-energy cosmic rays}

\author{Rafael \surname{Alves Batista}$^{1,2}$}
\email{rafael.alvesbatista@uam.es}
\affiliation{
	$^1$ Instituto de F\'isica Te\'orica UAM-CSIC, Universidad Aut\'onoma de Madrid, C/ Nicol\'as Cabrera 13-15, 28049 Madrid, Spain\\
	$^2$ Departamento de F\'isica Te\'orica, Universidad Aut\'onoma de Madrid, M-15, 28049 Madrid, Spain
}

\begin{abstract}
Recently an extraordinarily bright gamma-ray burst, \grb, was observed by several facilities covering the whole electromagnetic spectrum. Gamma rays with energies up to 18~TeV were detected, as well as a possible photon with 251~TeV. Such energetic events are not expected because they would be attenuated by pair-production interactions with the extragalactic background light. This tension is, however, only apparent, and does not call for any unconventional explanation. Here I show that these observations can be interpreted as the result of ultra-high-energy cosmic rays (UHECRs) interacting with cosmological radiation fields during their journey to Earth, provided that intergalactic magnetic fields are reasonably weak. If this hypothesis is correct, it would establish bursts like \grb as UHECR sources.
\end{abstract}



\maketitle

\section{Introduction}
\label{sec:intro}

On 9 October 2022 the Burst Alert Telescope (BAT) aboard the \textit{Swift} satellite observed an astonishingly bright gamma-ray burst (GRB), \grb~\cite{2022GCN32632}. This burst had triggered Fermi's Gamma-ray Burst Monitor (GBM) about one hour before~\cite{2022GCN32636}, but was only confirmed as a GRB later~\cite{2022GCN32642}. It was also detected by the Large Area Telescope (Fermi~LAT)~\cite{2022GCN32637}. This object is located at redshift $z = 0.151$, having an estimated isotropic-equivalent energy of $E_\text{tot} \simeq 2 \times 10^{47} \; \text{J}$, based on the fluence measured by Fermi~GBM~\cite{2022GCN32648}.
The Large High Altitude Air Shower Observatory (LHAASO) observed thousands of gamma rays from this same direction within $\sim 2000 \; \text{s}$ of the burst~\cite{2022GCN32677}. The observed photons have energies extending up to $E \simeq 18 \; \text{TeV}$. Interestingly, the array Carpet-2 at the Baksan Neutrino Observatory reported a photon-like air shower triggering the detectors within the same time window, with a (pre-trial) significance of $3.8\sigma$, which points to a primary photon of $E \simeq 251 \; \text{TeV}$~\cite{2022ATel15669}. 

Gamma rays from \grb with energies $E \gtrsim 10 \; \text{TeV}$ should be strongly suppressed due to pair production ($\gamma + \gammabg \rightarrow e^+ + e^-$) with backgrounds photons ($\gammabg$), mostly from the extragalactic background light (EBL). Notwithstanding EBL uncertainties, LHAASO and Carpet-2 would likely not observe \emph{primary} gamma rays from the burst, even for weaker EBL models. For this reason, non-conventional physics such as Lorentz invariance violation (LIV)~\cite{baktash2022a, li2022a, zhu2022a} and axion-like particles (ALPs)~\cite{baktash2022a, galanti2022a, troitsky2022a, lin2022a} have been quickly invoked to interpret the observations. A more conventional explanation would be the misidentification of a cosmic-ray shower by LHAASO as a genuine gamma ray detection~\cite{baktash2022a, zhao2022a} -- which is less likely if the Carpet-2 detection is true.

Here I suggest a plausible interpretation of these observations that does not require any unconventional physics. If the GRB emitted ultra-high-energy cosmic rays (UHECRs) with energies $E \gtrsim 1 \; \text{EeV}$ ($1 \; \text{EeV} \equiv 10^{18} \; \text{eV}$) towards Earth, then it is possible that they interacted with background photons, ultimately accounting for at least part of the observed signal. The necessary conditions for the validity of this hypothesis are the following:
\begin{enumerate}[label=(\roman*), noitemsep, topsep=2pt, leftmargin=8mm]
\setlength{\parskip}{2pt}
\item \label{enum:condition1} UHECRs should not be significantly deflected by magnetic fields before they produce the particles responsible for the observed gamma rays. 
\item \label{enum:condition2} Part of these gamma rays should arrive within the same time window as the observations, retaining temporal and angular correlation with the burst.
\item \label{enum:condition3} Neutrinos are also produced through the very same interactions that produce gamma rays. Therefore, this model has to be compared with observational neutrino limits~\cite{2022GCN32665}.
\end{enumerate}
Conditions~\ref{enum:condition1} and~\ref{enum:condition2} essentially depend on the properties of intergalactic magnetic fields (IGMFs), discussed in section~\ref{sec:interpretation}. Condition~\ref{enum:condition3} is carefully analysed using the numerical simulations described in section~\ref{sec:simulations}.

\medskip
As a prelude to this work, one must first establish that GRBs are suitable UHECR sources (for reviews, see refs.~\cite{alvesbatista2019d, anchordoqui2019a}). Indeed, they have long been deemed to be adequate UHE sources (e.g.,~\cite{waxman1995a, murase2006a, zhang2018a, biehl2018a, samuelsson2019a, boncioli2019a}). They satisfy the Hillas criterion~\cite{hillas1984a}, according to which charged particles of charge $q$ can be accelerated by shocks of velocity $v_\text{sh}$ up to energies $E_\text{max} \sim \eta^{-1} v_\text{sh} c q B R$, where $B$ and $R$ denote, respectively, the magnetic field and characteristic scale of the site wherein acceleration takes place, and $\eta$ is an efficiency factor. The actual details of how particle acceleration occurs are not trivial and vary depending on the properties of the burst. 
\grb is thought to be a collapsar~\cite{2022GCN32686}, i.e., the result of exceedingly powerful supernova. It is beyond the scope of this work to delve into the details of how CRs are accelerated in GRBs; for that, the reader is referred to refs.~\cite{murase2019a, meszaros2019a}.
At this stage, it suffices to know that many models posit that GRBs are sources of UHECRs. Once they escape the environment of the GRB, however that happens, they travel to Earth and their interactions with cosmological photons can lead to appreciable fluxes of cosmogenic particles along the line of sight, including neutrinos and photons. The details of how this occurs are described in section~\ref{sec:propagation}.

\section{Propagation of cosmic rays and gamma rays}\label{sec:propagation}

UHECRs can interact with cosmological photon fields such as the EBL, as well as the cosmic microwave background (CMB) and the cosmic radio background (CRB) during their journey to Earth. For a given cosmic-ray nucleus $\nucleus{A}{Z}$ of atomic number $Z$ and mass $A$, the main photonuclear processes that affect them are: photopion production (e.g., $p + \gammabg \rightarrow p + \pi^0$, $p + \gammabg \rightarrow p + \pi^+$), Bethe-Heitler pair production ($\nucleus{A}{Z} + \gammabg \rightarrow \nucleus{A}{Z} + e^+ + e^-$), and photodisintegration ($\nucleus{A}{Z} + \gammabg \rightarrow \nucleus{A-1}{Z} + n$, $\nucleus{A}{Z} + \gammabg \rightarrow \nucleus{A-1}{Z} + p$, etc). Unstable nuclei undergo alpha, beta, and gamma decays during the photodisintegration chain. An additional subdominant energy-loss channel for photon production is elastic scattering ($\nucleus{A}{Z} + \gammabg \rightarrow \nucleus{A}{Z} + \gamma)$. 
The by-products of these interactions are the still-undetected~\cite{alvesbatista2019a} cosmogenic particles, whose existence has been predicted long ago~\cite{berezinsky1969a, stecker1973b}. 

Electrons and photons, too, interact during intergalactic propagation. The main ones are pair production ($\gamma + \gammabg \rightarrow e^+ + e^-$) and inverse Compton scattering ($e^\pm + \gammabg \rightarrow e^\pm + \gamma$), although higher-order processes such as double pair production ($\gamma + \gammabg \rightarrow e^+ + e^- + e^+ + e^-$) and triplet pair production ($e^\pm + \gammabg \rightarrow e^\pm + e^+ + e^-$) may also contribute at some specific energy ranges. These processes feed one another, thus constituting an electromagnetic cascade (see ref.~\cite{alvesbatista2021a} for a detailed review). 

In addition to the aforementioned interactions, all particles lose energy losses due to the adiabatic expansion of the universe. Moreover, charged particles can emit synchrotron radiation in the presence of magnetic fields.

\bigskip
Cosmogenic particles can be produced approximately along the line of sight, leading to interesting observational signatures. Such model has been invoked, for example, to explain gamma-ray observations from blazars~\cite{essey2010a, essey2010b, essey2011a, kusenko2012a}, and also \grb~\cite{mirabal2022a, das2023a, rudolph2023a}. 
This directional correlation with sources can only occur if magnetic fields are not exceedingly strong. For a magnetic field of coherence length $L_B$, the deflection of a charged particle after travelling a distance $\ell$ is $\Delta \delta_B \simeq \arcsin R_\text{L} / \ell$ if $\ell \ll L_B$ and $\sqrt{2 \ell L_B} / 3R_\text{L}$ otherwise~\cite{lee1995a}, with $B$ denoting the strength of the magnetic field and $R_\text{L}$ the Larmor radius of the particle. The associated time delay is~\cite{alcock1978a}:
\begin{equation}
	\Delta t_B \simeq 
	\begin{cases}
		\dfrac{\ell}{c} \left[1 - \cos(\Delta\delta_B)\right] \;\; &\text{if} \;\; \ell \ll L_B \,, \\[10pt]
		\dfrac{\ell \Delta\delta_B ^2}{12 c} \;\; &\text{if} \;\; \ell \gg L_B  \,.
	\end{cases}
	\label{eq:timeDelay}
\end{equation}

The main source of uncertainty in $\Delta t_B$ are the properties of magnetic fields (see, e.g., refs.~\cite{vallee2011a, ryu2012a, hutschenreuter2018a} for reviews). UHECRs deflections cannot be properly estimated because IGMFs are unknown, especially in cosmic voids. Nevertheless, even in the most extreme scenarios they can be bounded~\cite{alvesbatista2017c}.

\section{Simulations}\label{sec:simulations}

To interpret the observations, one-dimensional Monte Carlo simulations are performed using the CRPropa code~\cite{alvesbatista2016a, alvesbatista2022a}. UHECRs are assumed to be emitted by the GRB with spectrum $\dd N / \dd E \propto E^{-\alpha}$, with an exponential suppression factor $\exp(-E / Z \mathcal{R}_\text{max})$ for $E \geq Z \mathcal{R}_\text{max}$. Here $\mathcal{R}_\text{max}$ denotes the maximal rigidity of the emitted cosmic rays, given by $\mathcal{R}_\text{max} \equiv E_\text{max} / Z$. A fraction $\eta_\text{CR}$ of the isotropic-equivalent energy of the GRB is assumed to be converted into cosmic rays (of all energies). To ensure that the secondary fluxes arrive within $\lesssim 1 \; \text{day}$ of the burst, magnetic deflections ought to be small. Since IGMFs are poorly know, a \emph{conservative} and an \emph{optimistic} scenario are considered, depending on the rigidity of the UHECRs. This established a lower bound on the rigidity of the primary UHECRs that will be considered. 

The interaction processes described in section~\ref{sec:propagation} are included in the analysis. As a benchmark for the simulations, the EBL model by Gilmore~\textit{et al.}~\cite{gilmore2012a} is chosen, as well as the CRB model by Protheroe \& Biermann~\cite{protheroe1996a}. Naturally, these choices have a considerable impact on the results of the simulation, together with other factors like photonuclear cross sections, for example~\cite{alvesbatista2015d, alvesbatista2019c}. It is beyond the scope of the present Letter to discuss these uncertainties in detail. It is already enough to prove that the proposed interpretation of \grb is justified considering at least one \emph{realistic} model.

\section{Analysis and Interpretation of the Observations}\label{sec:interpretation}

\grb drew much attention because the events detected by LHAASO and Carpet-2, in principle, should not have arrived at Earth due to pair production with the EBL. If \emph{primary} gamma rays were emitted, to first order, this flux ($\Phi_0$) would be exponentially suppressed by a factor corresponding to the optical depth ($\tau$):
\begin{equation}
	\Phi(E, z) = \Phi_0(E, z) \exp\left[-\tau(E, z)\right] \;	
	\label{eq:eblAttenuation}
\end{equation}
where $z$ is the redshift. It is understandable that a spectrum extending up to $\approx 18 \; \text{TeV}$ from a source at $z \simeq 0.15$ could be seen as a signature of New Physics (see ref.~\cite{addazi2022a} for a discussion of some models), especially with a possible coincident event with $E \approx 251 \; \text{TeV}$. But many plausible hypotheses remain within the realm of conventional explanations. An UHECR origin is one of them, provided that the requirements \ref{enum:condition1}, \ref{enum:condition2}, \ref{enum:condition3} are fulfilled.

\medskip
To obtain a rough estimate for the time delay of photons from \grb for the scenario here proposed, it is important to know its precise location within the large-scale structure of the universe. The contribution of the host galaxy of the GRB can be ignored given the jet's extension. For this reason, it is a good approximation to consider only the contributions of galaxy clusters, filaments, and cosmic voids. The total time delay due to the magnetic fields in each of these regions is:
\begin{equation}
	\Delta t_B \simeq \Delta t_{B, \text{cluster}} + \Delta t_{B, \text{filament}} + \Delta t_{B, \text{void}} \,.
\end{equation}
Here $\Delta t_{B, \text{cluster}}$ and $\Delta t_{B, \text{filament}}$ refer to the time delays associated to the specific cluster and filament where the GRB is embedded. To first order, it is reasonable to neglect all crossings of the particles with external clusters and filaments given that they respectively amount to $\lesssim 10^{-4}$ and $\lesssim 0.2$ of the universe's volume. 

Galaxy clusters have magnetic fields around $10^{-13} \lesssim B_\text{cluster} / \text{T} \lesssim 10^{-10}$~\cite{ryu2012a, vazza2017a}. Lower-mass clusters are much more abundant than massive ones. The magnetic field near their centre is comparatively lower by a ten fold. They typically have a size of $\sim 1 \; \text{Mpc}$ with a radially decreasing field. Therefore, time delays for particles with rigidities of $\mathcal{R} \sim 100 \; \text{EV}$ would be $\Delta t_{B, \text{cluster}}  \lesssim 10^{11} \; \text{s}$, according to eq.~\ref{eq:timeDelay}. This is an upper limit. \grb was more likely located in an average-sized cluster since they are more abundant, possibly outside the inner region or in the outskirts. With these realistic considerations, $\Delta t_{B, \text{cluster}}  \lesssim 10^{5} \; \text{s}$, which in more optimistic scenarios could go down by up two orders of magnitude.

Magnetic fields in the filaments connecting galaxy clusters are uncertain, but they are believed to be $10^{-15} \lesssim B / \text{T} \lesssim 10^{-13}$~\cite{vazza2017a, gheller2019a, osullivan2020a}. If the geometry of the emission was such that the GRB jet was aligned with a filament,  a time delay would be expected over a distance of the order of its length, yielding $\Delta t_{B, \text{filament}} \sim 10^9 \; \text{s}$ for $\mathcal{R} = 100 \; \text{EV}$. This is not a very probable scenario considering that to each cluster there are only a handful of associated filaments, whose traverse widths are comparable to the cluster size. Therefore, the probability of the jet to be perfectly aligned with a cluster is small. This implies that the emitted particles would travel only a few Mpc through the filament, resulting in $\Delta t_{B, \text{filament}} \sim 10 \; \text{s}$.

Cosmic voids fill most of the universe's volume, up to 80\%, but their magnetic fields are unknown, ranging between $10^{-15} \; \text{T}$~\cite{jedamzik2019a} and $10^{-20} \; \text{T}$~\cite{fermi2018a}. They incur time delays of $10^{2} \lesssim \Delta t_{B, \text{void}} / \text{s} \lesssim 10^{8}$ over a distance $\ell$ comparable to the mean free of 100~EV particles, which is typically $\lambda_\text{CR} \simeq 100 \; \text{Mpc}$ for UHECRs. Therefore, a realistic upper bound on the total time delay of the UHECRs emitted by the GRB would be $\Delta t_B \lesssim 10 ^ {8} \; \text{s}$. The main source of uncertainty remains the exact location of the GRB progenitor within its host galaxy cluster, which could shrink this estimate down to $\Delta t_B \lesssim 10 ^ {4} \; \text{s}$. 

The charged component of electromagnetic cascades can be substantially delayed. The dominant contribution in this case comes from the fields in the voids. Electrons with $E \lesssim 1 \; \text{PeV}$ have mean free paths $\lambda_\text{e} \sim 10 \; \text{kpc}$, which for $B \lesssim 10^{-16} \; \text{T}$ results in $\Delta t_B \lesssim 10 ^ {5} \; \text{s}$ for electrons with $E \gtrsim 1 \; \text{PeV}$, conservatively assuming $\lambda_\text{e} \ll L_B$. These are electrons that could potentially generate energetic gamma rays with $E \sim 100 \; \text{TeV}$. 

These considerations satisfy conditions~\ref{enum:condition1} and \ref{enum:condition2} described in section~\ref{sec:intro}. To address~\ref{enum:condition3}, the numerical simulations described in section~\ref{sec:simulations} must be analysed.

\begin{figure*}
	\centering
	\includegraphics[width=\columnwidth]{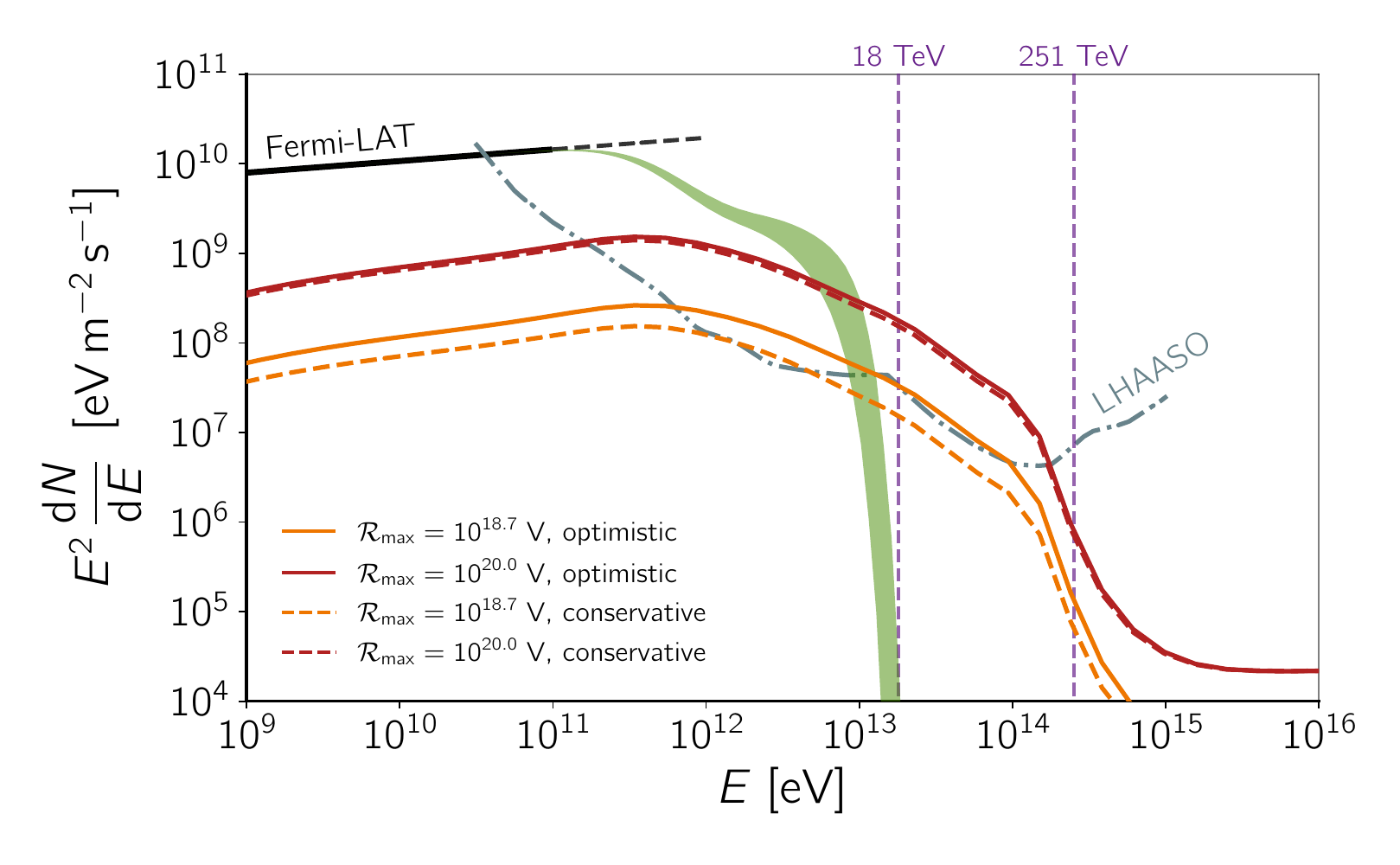}
	\includegraphics[width=\columnwidth]{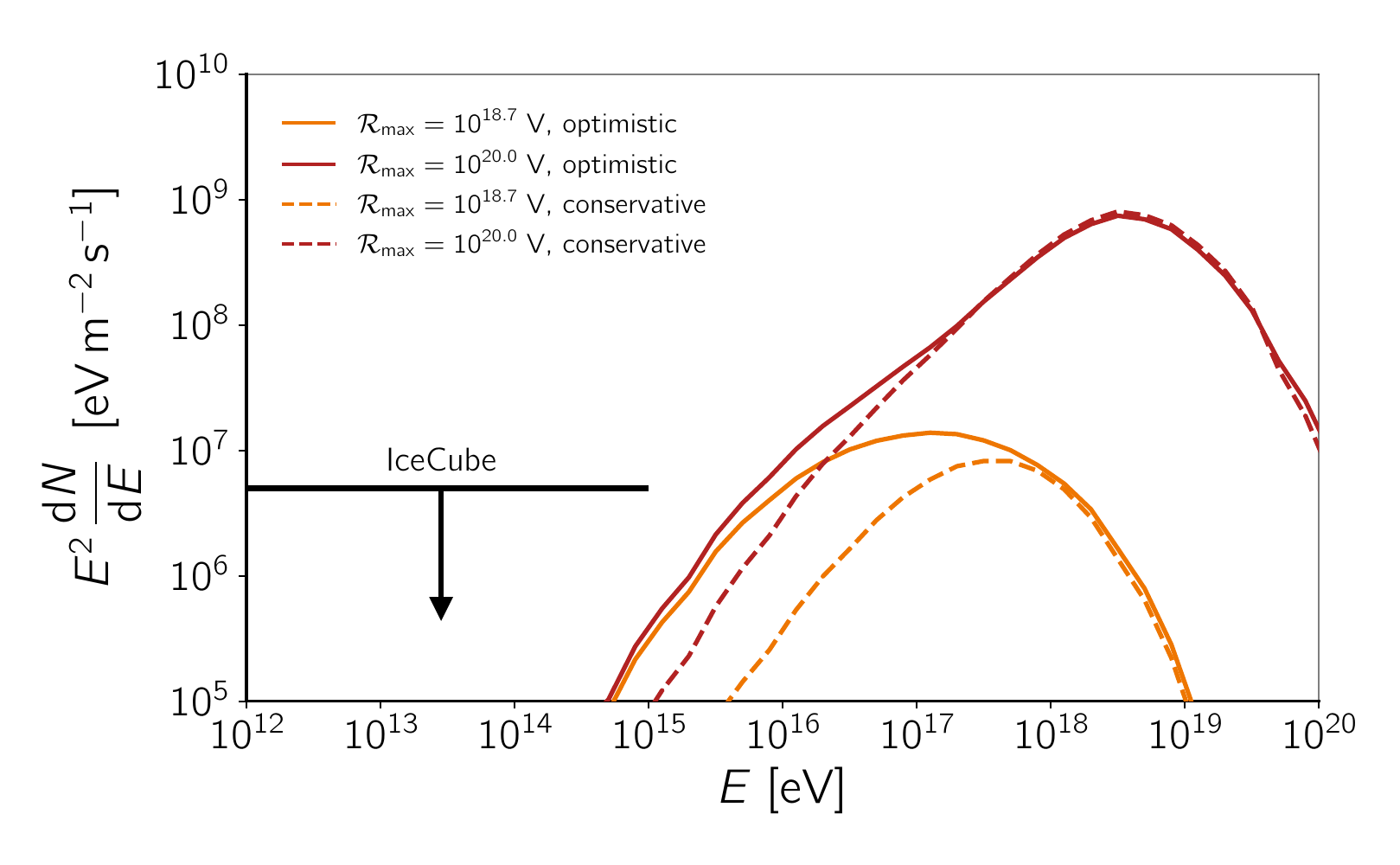}
	\includegraphics[width=\columnwidth]{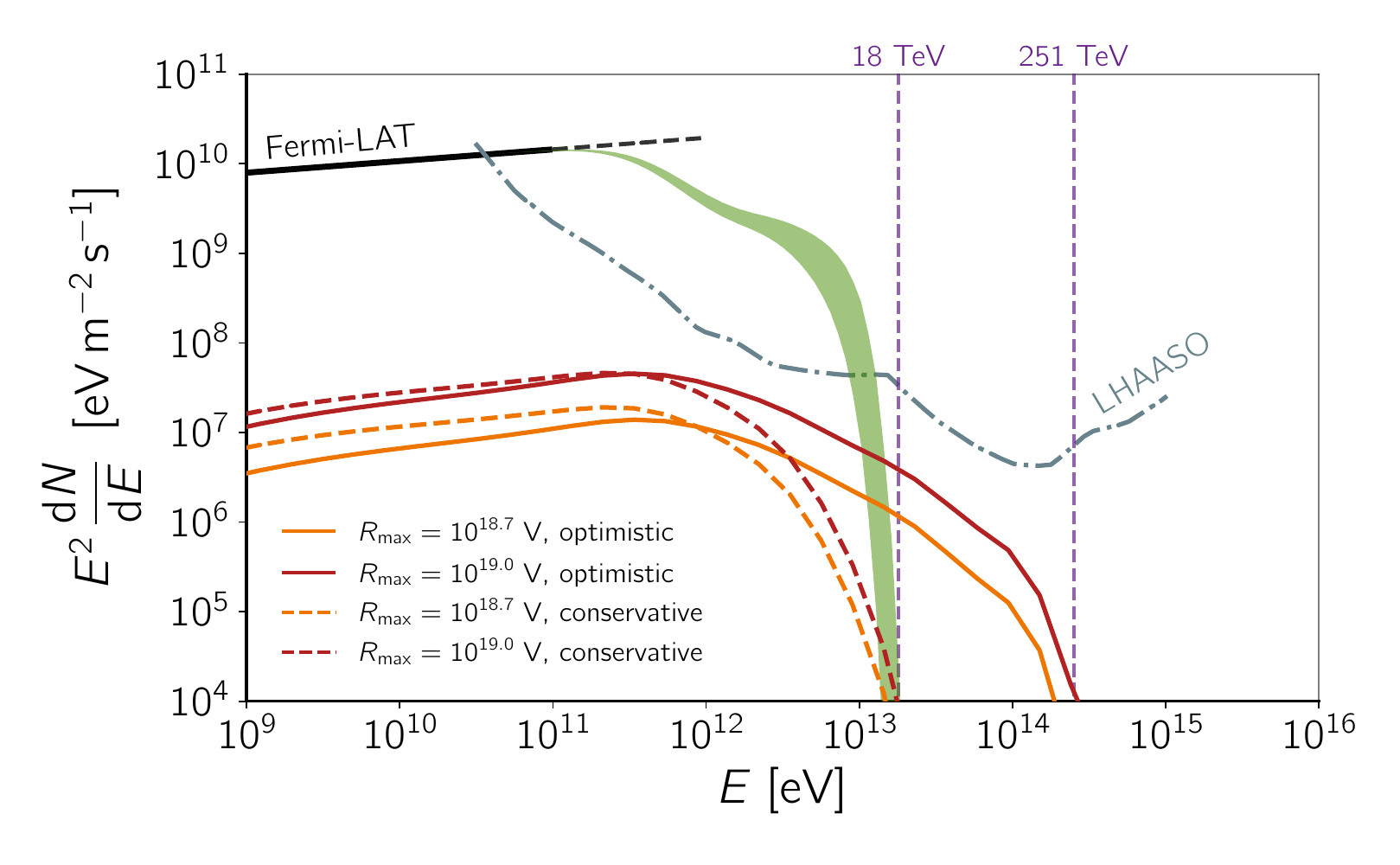}
	\includegraphics[width=\columnwidth]{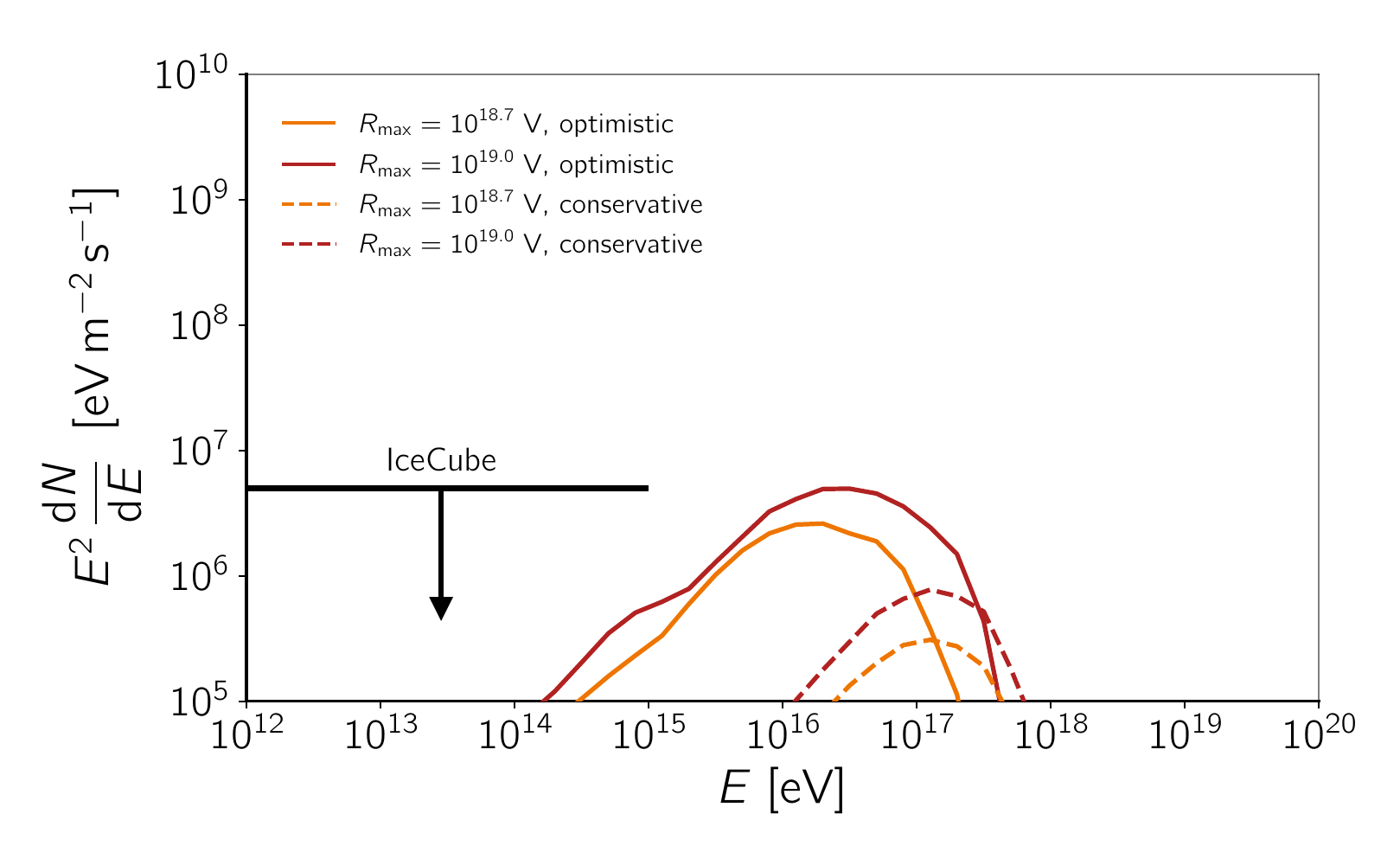}
	\caption{This figure illustrates the flux of cosmogenic photons (left) and all flavours of neutrinos (right column) assuming protons (upper panels) and nitrogen primaries (lower panels). The \emph{conservative scenario} assumes strong magnetic deflections, and hence only UHECRs with $\mathcal{R} \gtrsim 30 \; \text{EV}$ are considered, whereas the \emph{optimistic scenario} corresponds to weaker fields, thus $\mathcal{R} \gtrsim 1 \; \text{EV}$. The spectral index is assumed to be $\alpha = 2.0$ for the maximal rigidities ($\mathcal{R}_\text{max}$) indicated in the legend, assuming that 0.5\% of the total luminosity is converted into CRs. The vertical dashed lines indicate the maximum energy of the photons detected by LHAASO~\cite{2022GCN32677} and Carpet-2~\cite{2022ATel15669}. The green band represents the expected flux of primary gamma rays reaching Earth obtained from eq.~\ref{eq:eblAttenuation} and including EBL uncertainties, according to the model by Domínguez~\textit{et al.}~\cite{dominguez2011a}. Measurements by Fermi-LAT~\cite{2022ATel15656} are also shown, as well as limits by IceCube~\cite{2022GCN32665}.}
	\label{fig:cosmogenic}
\end{figure*}

\medskip
The results of the simulations of UHECR and gamma-ray propagation are shown in fig.~\ref{fig:cosmogenic}. The left panels correspond to the predicted photon fluxes, whereas the ones on the right depict the estimated neutrino fluxes. Results for primary UHECR protons and nitrogen are shown.

Fermi-LAT~\cite{2022ATel15656} inferred a flux of $(62 \pm 4) \; \text{photons} \, \text{m}^{-2} \, \text{s}^{-1}$, with a photon index of $\alpha = 1.87 \pm 0.04$. These results were obtained considering the energy range 0.1--1~GeV, between 200~s and 800~s after the GBM trigger. The highest-energy photon observed had 99~GeV.  This is shown on the left panels of fig.~\ref{fig:cosmogenic}, together with the extrapolation of these measurements to higher energies. Furthermore, the High-Altitude Water Cherenkov observatory (HAWC) provided a 95\% C.L. upper limit on the differential energy flux assuming an $E^{-2}$ spectrum, $\simeq 4.2 \times 10^{16} \; \text{eV} \, \text{m}^{-2} \, \text{s}^{-1}$ ~\cite{2022GCN32683}, also indicated. 
The fact that in this figure the gamma-ray fluxes predicted by the simulations extend up to energies far higher than the theoretical expectation for the primary gamma-ray scenario attests to the plausibility of this scenario. Moreover, depending on the spectrum of the emitted UHECRs the photon fluxes can be nearly as high as Fermi-LAT's measurements. 

The IceCube Neutrino Observatory searched for track-like muon neutrino events, under the assumption of a $E^{-2}$ spectrum. The 90\% C.L. flux upper limits are $3.9 \times 10^{11} \; \text{eV} \, \text{m}^{-2}$ for a search starting 1~hour before the GBM trigger up to 2~hours thereafter. This upper limit, shown in fig.~\ref{fig:cosmogenic}, weakens if a time window of $\pm$~1~day is considered, going up to $4.1 \times 10^{11} \; \text{eV} \, \text{m}^{-2}$~\cite{2022GCN32665}. The Cubic Kilometre Neutrino Telescope (KM3NeT) repeated similar analysis, also finding no events, despite the sub-optimal location of the GRB within its field of view at the time of the burst~\cite{2022GCN32741}.

\section{Discussion}\label{sec:discussion}

The predicted fluxes from fig.~\ref{fig:cosmogenic} are exclusively for a cosmogenic origin. In reality, a fraction of the observations could likely be due to primary gamma rays, especially for $E \lesssim 10 \; \text{TeV}$. Disentangling these two components is no easy task, especially in the absence of a neutrino counterpart.

So far there have been no reports of UHE neutrino limits. For the declination of \grb, the neutrino point-source sensitivity of the Pierre Auger Observatory lies between $\sim 5 \times 10^{6}$ and  $3 \times 10^{7} \; \text{eV} \, \text{m}^{2} \, \text{s}^{-1}$, for a single neutrino flavour, for energies between 100~PeV and 100~EeV~\cite{auger2012e, auger2019b}. This sensitivity probes part of the parameter space, as seen in fig.~\ref{fig:cosmogenic}. A dedicated analysis by Auger could deliver significant constraints on GRB properties at ultra-high energies.

The IceCube Collaboration performed a search for correlations between the arrival direction of high-energy neutrinos in coincidence with the prompt emission of nearly 2000~GRBs. They found no significant correlations~\cite{icecube2017d}. This places constraints on the total fluxes of neutrinos, both from the source and cosmogenic, and on UHECR production in GRBs. The latter component, however, requires a broader time window for the analysis to accommodate time delays.

Realistically, at least part of the UHECR flux from the GRB would not reach Earth within an anthropic time window of a few decades, so it serves no purpose to address this condition at the moment. Nonetheless, it is reasonable to suppose that past events analogous to \grb can contribute to the UHECR flux measured today. 

The exact location of \grb in the universe is arguably the most relevant parameter to support the hypothesis being put forth here. Ref.~\cite{mirabal2022a} stated that the explosion very likely happened in a void. While this could be true, it lacks supporting evidence. The probability that \grb exploded somewhere within denser region is overwhelmingly higher, as discussed in  section~\ref{sec:interpretation}. Any reliable interpretation of the observations \emph{necessarily} requires considerations about particle propagation in clusters and filaments. This justifies the inclusion of a lower-energy rigidity-dependent cut-off and the analysis of both a conservative and an optimistic scenarios (see fig.~\ref{fig:cosmogenic}). 

Future facilities capable of detecting UHE photons and UHE neutrinos could help test the hypothesis~\cite{ackermann2022b}. For instance, the next-generation IceCube detector, IceCube-Gen2~\cite{icecubegen22021a}, the Probe of Extreme Multi-Messenger Astrophysics (POEMMA)~\cite{poemma2021a}, and the Giant Radio Array for Neutrino Detection (GRAND)~\cite{grand2020a} will have enough sensitivity at $E \gtrsim 100 \; \text{PeV}$ to fully investigate the plausibility of this scenario. The non-detection of neutrinos in this energy band would exclude the an UHECR origin for the highest-energy gamma-ray events observed.

It is important to bear in mind that other explanations for the energetic events observed by Carpet-2 and LHAASO in coincidence with \grb are also possible. The location of the event -- near the Galactic plane -- begs for a careful statistical analysis to determine the likelihood of putative Galactic sources to emit gamma rays with energies $\gtrsim 100 \; \text{TeV}$ in angular and temporal coincidence with \grb, as noted in ref.~\cite{2022ATel15675}. One should also take into account the fact that that these are air-shower observatories, such that they have poor angular resolution compared with the other facilities observing the GRB at the time, thereby increasing the chance of random coincidences. 

\section{Summary \& Outlook}

The analysis here presented demonstrates that gamma rays with energies of up to $\sim 1 \; \text{PeV}$ from \grb could reach Earth if they are by-products of UHECR interactions during propagation. No phenomena such as LIV or photon-ALP interconversion are required to explain the observations.
The real question is whether temporal and angular coincidences would be preserved in this case, considering that both the UHECRs and the charged component of the electromagnetic cascades could be delayed with respect to the burst. Through simple estimates and one-dimensional simulations I argued that the highest-energy cosmic rays with $E \gtrsim 40 \; \text{EeV}$ could realistically evade this limitation if: IGMFs in cosmic voids are weaker than $\lesssim 10 ^ {-17} \; \text{T}$; the GRB is not located in the central region of its host galaxy cluster; the GRB jet is not aligned with cosmic filaments. All of these conditions can be satisfied realistically.  
Therefore, \grb could be a source of UHECRs.

\section*{Acknowledgements}

This work was funded by the ``la Caixa'' Foundation (ID 100010434) and the European Union's Horizon~2020 research and innovation program under the Marie Skłodowska-Curie grant agreement No 847648, fellowship code LCF/BQ/PI21/11830030. It was also supported by the grants PID2021-125331NB-I00 and CEX2020-001007-S, both funded by MCIN/AEI/10.13039/501100011033 and by ``ERDF A way of making Europe''.

%%%%%%%%%%%%%%%%%%%%%%%%%%%%%%%%%%%%%%%%%%%%%%%%%%
%%%%%%%%%%%%%%%%%%%% REFERENCES %%%%%%%%%%%%%%%%%%

\bibliographystyle{apsrev4-1}
\bibliography{/Users/rab/Dropbox/library/references, /Users/rab/Dropbox/library/telegrams}

\end{document}